\documentclass[namedreferences]{kluwer}

\usepackage{graphicx}

\begin{document}
\begin{article}

\begin{opening}
\title{Procedures, Resources and Selected Results of the Deep Ecliptic Survey}

\author{M. W. \surname{Buie}\email{buie@lowell.edu}}
\author{R. L. \surname{Millis}\email{rlm@lowell.edu}}
\author{L. H. \surname{Wasserman}\email{lhw@lowell.edu}}
\institute{Lowell Observatory}
\author{J. L. \surname{Elliot}\email{jle@mit.edu}}
\institute{Massachusetts Institute of Technology and Lowell Observatory}
\author{S. D. \surname{Kern}\email{susank@mit.edu}}
\author{K. B. \surname{Clancy}\email{kclancy@mit.edu}}
\institute{Massachusetts Institute of Technology}
\author{E. I. \surname{Chiang}\email{echiang@astron.berkeley.edu}}
\author{A. B. \surname{Jordan}\email{abjordan@uclink4.berkeley.edu}}
\institute{University of California at Berkeley}
\author{K. J. \surname{Meech}\email{meech@ifa.hawaii.edu}}
\institute{University of Hawaii}
\author{R. M. \surname{Wagner}\email{rmw@as.arizona.edu}}
\institute{Large Binocular Telescope Observatory}
\author{D. E. \surname{Trilling}\email{trilling@astro.upenn.edu}}
\institute{University of Pennsylvania}
\runningtitle{Deep Ecliptic Survey}
\runningauthor{Buie et al.}
\begin{ao}
\\
Marc W. Buie\\
Lowell Observatory\\
1400 W. Mars Hill Rd.\\
Flagstaff, AZ~~86001~~USA\\
email: buie@lowell.edu
\end{ao}

\begin{abstract}

The Deep Ecliptic Survey is a project whose goal is to survey a large
area of the near-ecliptic region to a faint limiting magnitude ($R \sim
24$) in search of objects in the outer solar system.  We are collecting
a large homogeneous data sample from the Kitt Peak Mayall 4-m and Cerro
Tololo Blanco 4-m telescopes with the Mosaic prime-focus CCD cameras.
Our goal is to collect a sample of 500 objects with good orbits to further
our understanding of the dynamical structure of the outer solar system.
This survey has been in progress since 1998 and is responsible for 272
designated discoveries as of March 2003.  We summarize our techniques,
highlight recent results, and describe publically available resources.

\end{abstract}
\end{opening}


\section{Summary and Goals of the Survey}

With the discovery of a large population of objects at or beyond the
distance of Neptune comes the ability to learn about the processes that
shape our solar system at much greater distances than previously possible.
Many groups are focusing on studying the newly discovered objects
themselves through color or spectroscopic measurements, lightcurve and
rotation studies, or searches for satellites.  However, before these
studies can proceed, one must first locate these objects and determine
good orbits.  Over and above determining existence, having an ensemble
of good orbits permits investigations into the dynamical structure
of the outer solar system.  We wish to understand which objects are
likely to be dynamically long-lived and thus may represent undisturbed
primordial material.  We wish to understand dynamical lifetimes and the
role resonances have played over the age of the solar system.

Our project began in 1998 as the first of the large-format CCD mosaic
array cameras was commissioned by NOAO at Kitt Peak.  Utilizing the
Mosaic cameras (0.35 arcmin$^2$, 8k$\times$8k array) at Kitt Peak
and Cerro Tololo, we have been pursuing a systematic survey of the
near-ecliptic region.  In 2001 we were granted 3 years of formal survey
status by NOAO which provides 20 nights of access per year (now extended
to a fourth year).  The goal of the survey is to collect a sample of
500 KBOs with well-established orbits and known observational biases
in their discoveries.  From this dataset we can address questions of
resonance population, inclination distribution, radial distribution,
magnitude or size distribution, and more.  In this article we present a
brief summary of our survey technique, provide a snapshot of our results,
and describe resources we have developed for our survey that may be of
wider interest to the scientific community.

\section{Summary of Techniques}
\subsection{Data Collection}

We have divided the sky up into 12600 field locations whose field centers
are within 6.5 degrees of the ecliptic.  The abutting fields correspond
to the area of the Mosaic array on the sky.  Any field (collection of
8 detector images) with one or more stars brighter than $R=9$ is not
used for searching.  Also, only those fields that contain 35 or more
USNO-A2.0 catalog sources on each CCD are then used for searching.
This list contains 1943 total fields at fixed locations that are
distributed at all ecliptic longitudes, except for a 20$^{\circ}$ gap
where the ecliptic crosses the galactic plane near the galactic center.

Candidate fields for any given observing run lie within 30$^{\circ}$
of the opposition point, thus eliminating confusion with the far more
numerous main-belt asteroids.  We search a field by taking two images on
one night separated by between 1.7 and 2.7 hours.  We usually attempt a
third frame for confirmation on either the preceding or following night.
In the early days of the survey, this third frame was not regularly
obtained.  Foregoing the third frame enabled us to maximize sky coverage
and, therefore, the number of objects discovered per night.  However,
without an observation on a second night, the Minor Planet Center will
not announce a discovery to the community.  Because this annoucement
was judged to significantly increase the likelihood of astrometric
observations by others, we began taking third frames of all fields
observed in a given run.

All images are binned 2x2 to reduce the readout time and the file size.
This produces a final image scale of 0.52 arcsec/pixel which is adequate
to sample the seeing disk on all but the best nights.  We rarely see
signs that the data are undersampled even though sub-arc second seeing
is common at these facilities.  This situation may well be due to the
broad VR filter ($\lambda_{{\rm cent}} = 6080$\AA, $\lambda_{{\rm FWHM}}
= 2230$\AA) that we use for maximum sensitivity, in that the atmospheric
dispersion correction is designed to correct a somewhat narrower bandpass.

\subsection{Data Processing}

Our data processing pipeline applies simple image processing rules.
We subtract overscan and correct for a superbias image on a chip-by-chip
basis.  Superbias images are generally created from a stack of 10 bias
images.  Flat fielding comes from taking the first 20-30 images of the
night --- typically the first set that is collected without a repeated
telescope pointing.  The data images are combined with sigma-outlier
exclusion with a mean of the pixels that do not include sources or
cosmic ray hits.  While these flats may not represent the best possible
calibration products, they are more than adequate to support our source
detection procedures.

After the data are flattened, we run a source detection program and
generate a list of all sources that are 3-sigma above sky or brighter.
Once located, we generate instrumental magnitudes for all sources with
the aperture size chosen for that night.  Generally we use an aperture
of 3 or 4 pixel radius, sometimes as large as 5 on poorer nights.

Following source detection, we solve for an astrometric transformation
from pixel coordinates to equatorial celestial coordinates using reference
stars from the USNO A2.0 star catalog.  The images are strongly distorted,
particularly in the corner CCDs, and sometimes determining an astrometric
solution can be challenging.  All astrometric solutions are visually
inspected and verified.  A last stage by-product of the astrometric
solution is a correspondence between instrumental magnitude and catalog
magnitude.  We select those stars with consistent PSFs to determine
a zero-point correction for each CCD.  The process of generating a
zero-point correction works very well but the final magnitudes are only
as good as the catalog (generally to a few tenths of a magnitude but
sometimes much worse, particularly at southern declinations).  Work is
underway to provide a firm photometric calibration of the survey data.

Given a list of sources with right ascension, declination, and magnitudes
for each frame pair, we then compare the two lists and remove all
common sources.  On the objects left over we look for pairs of objects
moving within 30 degrees of the ecliptic with motion rates at or below
5 arcsec/hour.  Any such pairs found are tagged for visual confirmation.

The next step in the process is a visual inspection of all image pairs to
search for moving objects.  We have an IDL-based program that displays
the images and allows marking of moving objects.  The image pairs are
shown with the first epoch in the red image plane and the second epoch
in the blue and green image planes.  Fixed objects (stars, galaxies,
etc.) show as white objects.  Moving objects appear as red/cyan pairs
that are very easy to spot.  This technique is much, much faster
than traditional blinking and quickly to identify moving objects.
Each image pair is examined successively by two people to minimize the
effects of visual fatigue.  The first pass is also used to validate any
objects found via software.  We have tested this method against software
detection algorithms and with data salted with synthetic moving objects
\cite{mil02}.  It compares very favorably against all-computer techniques,
with essentially no difference in limiting magnitude or completeness
except that we can reliably work with pairs of detection images where
software-only techniques need three or more frames to be effective.
The 50\% efficiency detection threshold is quite close to the limiting
magnitude of an image, which defined to be the magnitude of a source whose
peak pixel is 3-$\sigma$ above sky.  Visual inspection does, however,
have completely different detection biases from computer techniques,
so the final list of objects is more complete than could be obtained
from just one technique.  Note that in our survey we attempt to mark,
measure, and report all moving targets, regardless of rate.  Our field
size and cadence is such that we are sensitive to motion rates from 0.5
arcsec/hour up to about 200 arcsec/hour.

Once the images have been examined, we compute the right ascension,
declination and magnitudes of the moving objects from the previously
determined solutions.  One person then reviews all slowly moving
objects to make a final determination of the validity of the detection and
to consult the third image for confirmation (if it exists).

A final review of the candidate objects comes from attempting to fit a
family of V\"ais\"al\"a orbits to the observations.  If the resulting
orbit does not yield a meaningful prograde Kuiper-Belt or Centaur orbit,
the object is dropped from the ``interesting'' object list.  The most
common reason for rejection in this review is that the positions of the
candidate cannot be fit by any physically plausible orbit.

Once these reviews are completed, we compare all measurements against
known KBOs and Centaurs to find linkages with already known objects.
Once these linkages are identified, all observations are reported to the
Minor Planet Center.  Generally speaking, completion of the astrometric
reference frame determination takes no more than a day or two after
the end of the run.  In fact, if everything goes well during a run, it
is possible to leave the telescope with all the astrometric reductions
having been completed.  We can often start the visual inspection before
the end of the run as well.  The rest of the processing can be done in
as little as a week following the observing run provided that there are
no complications in the data.  Normally, the results are available and
reported within two to three weeks.

\section{Summary of Objects Found}

Including all overhead we can collect about 110 images per 10 hour night with
a 4-minute exposure time.
With 3 images per field, this amount yields 37 search fields, or 12.8
square-degrees per night.  As of 2003 Mar 1, we have been allocated 400
hours at Kitt Peak where 244 hours were usable (61\%) and 250 hours
at Cerro Tololo where 229 hours were usable (92\%).  To date we have
thus surveyed almost 600 square-degrees.  Over this region, we have
identified 468 interesting objects (moving slower than 15 arcsec/hour
within 30 degrees of the opposition point) of which 272 have received
official designations.

\subsection{Special List of ``Interesting'' Objects}

As the sample size of known KBOs increases, better statistics are
realized for mapping out previously known structures (eg., non-resonant
vs. resonant populations).  In addition to the routine objects found,
we have begun to find the more unusual objects in the outer solar system.
In Tables I-IV we provide shorts lists of particularly interesting
categories of objects that have well-determined orbits.  In each table
the object name is given along with the absolute magnitude, H$_{\rm V}$,
semi-major axis, $a$ (in AU), eccentricity $e$, inclination $i$, and
perihelion distance, $q$ (in AU).  The next column is the uncertainty
of the most relevant quantity in each table.  Finally, the aphelion
distance, $Q$ (in AU), is given followed by the dynamical classification
(discussed in the next section).  Note that the orbital elements listed and
their errors are determined by the method of \inlinecite{ber00} except that we
determine barycentric elements rather than heliocentric elements.  Also, the
elements are all integrated to a common epoch of 2003 Aug 5.

Table~\ref{tbl1} shows a list of the intrinsically brightest objects
(largest if a constant albedo is assumed).  Although this is a short list
of objects, there are a wide range of dynamical classes represented here.

\begin{table}
\caption{Intrinsically bright objects, H$_{\rm V}<5$}\label{tbl1}
\begin{tabular*}{\maxfloatwidth}{lcccrclcl} \hline
Object Name& H$_{\rm V}$& $a$& $e$&
\multicolumn{1}{c}{$i$}& $q$&
\multicolumn{1}{c}{$\sigma_q$}& $Q$& Type \\ \hline
(28978) Ixion&            3.2& 39.4&  0.243&  19.7& 29.8& 0.006& 49&3:2$e$,
6:4$i^2$ \\
(42301) 2001 UR$_{163}$&  4.2& 51.4&  0.281&   0.8& 37.0& 0.004& 66&ScatExtd
     \\
2001 QF$_{298}$&          4.6& 39.2&  0.113&  22.4& 34.8& 0.089& 43&3:2$e$
     \\
2001 UQ$_{18}$&           4.9& 44.2&  0.028&   5.2& 43.0& 0.45 & 45&Classical
     \\
(19521) Chaos&            4.9& 45.9&  0.108&  12.0& 41.0& 0.007& 51&Classical
     \\
2001 KA$_{77}$&           4.9& 47.4&  0.095&  11.9& 42.9& 0.053& 52&Classical
     \\
2000 CN$_{105}$&          4.9& 44.8&  0.095&   3.4& 40.6& 0.055& 49&Classical
     \\
\hline
\end{tabular*}
\end{table}

Table~\ref{tbl2} shows a list of the most distant objects: large
perihelion distances and large semi-major axes or large eccentricity.
The object 2000~OO$_{67}$
has the largest aphelion distance (by a wide margin) of any KBO or
Centaur discovered thus far.  One object of particular note here (2000
CR$_{105}$) has a perihelion distance well outside the orbit of Neptune
(see discussion of \opencite{gla02} and \opencite{bru02}).  Though its
orbit indicates a past scattering event, there are as yet no widely
accepted explanations for its origin.  Note that all of these objects
have moderately large inclinations.

\begin{table}
\caption{Distant objects, ($q>30$, $a>90$) or ($e>0.8$)}\label{tbl2}
\begin{tabular*}{\maxfloatwidth}{lcrccllcl} \hline
Object Name& H$_{\rm V}$&
\multicolumn{1}{c}{$a$}& $e$& $i$& \multicolumn{1}{c}{$q$}&
\multicolumn{1}{c}{$\sigma_q$}& $Q$& Type \\ \hline
2000 OO$_{67}$ & 9.1& 523.7& 0.960& 20.1& 20.8 & 0.05& 1027& Centaur\\
2001 FP$_{185}$& 6.1& 225.4& 0.848& 30.8& 34.25& 0.03&  416& ScatNear\\
2000 OM$_{67}$ & 6.3&  97.3& 0.597& 23.4& 39.19& 0.03&  155& ScatNear\\
2000 CR$_{105}$& 6.1& 229.8& 0.807& 22.7& 44.2 & 0.2 &  415& ScatExtd\\
\hline
\end{tabular*}
\end{table}

Table~\ref{tbl4} shows a list of the highest inclination objects from
our survey.  The last object on the table, 2002~XU$_{93}$, has the
largest inclination of any object found so far.  Only one other object,
2002~VQ$_{94}$ (not found in our survey) has a similar inclination, and
both are Centaurs.  In the non-Centaur population (including scattered
disk), inclinations greater than 35 degrees have not been seen.

\begin{table}
\caption{High-inclination objects, $i>30$}\label{tbl4}
\begin{tabular*}{\maxfloatwidth}{lcrccllrl} \hline
Object Name& H$_{\rm V}$&
\multicolumn{1}{c}{$a$}& $e$& $i$& \multicolumn{1}{c}{$q$}&
\multicolumn{1}{c}{$\sigma_i$}&
\multicolumn{1}{c}{$Q$}& Type \\ \hline
2001 QC$_{298}$& 5.9&  46.1& 0.120& 30.6& 40.6& 0.001 &  52& ScatNear\\
2001 FP$_{185}$& 6.1& 225.4& 0.848& 30.8& 34.3& 0.0001& 416& ScatNear\\
2000 QM$_{252}$& 6.8&  40.6& 0.071& 34.8& 37.7& 0.006 &  44& ScatNear\\
2002 XU$_{93}$ & 7.9&  68.5& 0.695& 77.8& 20.9& 0.058 & 116& Centaur\\
\hline
\end{tabular*}
\end{table}

Table~\ref{tbl5} includes all ``nearby'' objects from our survey.
These objects have orbits at or interior to Neptune but beyond Jupiter.
Most of these objects are Centaurs but also included is the first known
Neptune Trojan object \cite{chi03a}.

\begin{table}
\caption{Objects with aphelion at or inside Neptune's orbit,
$Q<35$}\label{tbl5}
\begin{tabular*}{\maxfloatwidth}{lcccrllrl} \hline
Object Name& H$_{\rm V}$& $a$& $e$&
\multicolumn{1}{c}{$i$}& \multicolumn{1}{c}{$q$}&
\multicolumn{1}{c}{$\sigma_Q$}&
\multicolumn{1}{c}{$Q$}& Type \\ \hline
(54598) 2000 QC$_{243}$&  7.6& 16.5& 0.202& 20.8& 13.1& 0.0007 & 19.8&
Centaur\\
2000 CO$_{104}$        & 10.0& 24.3& 0.152&  3.1& 20.6& 0.009  & 27.9&
Centaur\\
2002 PQ$_{152}$        &  8.5& 25.6& 0.196&  9.4& 20.6& 0.2    & 30.6&
Centaur\\
2001 QR$_{322}$        &  7.0& 30.1& 0.017&  1.3& 29.6& 0.05   & 30.6& 1:1
\\
2001 KF$_{77}$         &  9.4& 26.0& 0.238&  4.4& 19.8& 0.009  & 32.2&
Centaur\\
\hline
\end{tabular*}
\end{table}

\subsection{Orbit Classifications}

With the increasing number of objects and the emergence of new dynamical
complexities in this sample, we have been forced to revisit the concept
of orbit classification.  Previous schemes are inadequate to capture
the dynamical diversity and similarities in the trans-Neptunian region.
Also, attributing membership to a particular resonance has usually relied
on an assignment based on semi-major axis, eccentricity, and inclination.
Our improved process is discussed at some length in \inlinecite{chi03b}
and uses a 3~My forward integration upon which the orbit classification
is based.  Table~\ref{tbl6} lists the dynamical type for all objects
from our survey with observation arcs greater than 30 days.
`Qual'' is a orbit classification quality code
where 3 means that the nominal orbit and its $\pm 3 \sigma$ clones agree,
2 means that one clone matches the nominal orbit, 1 means that both
clones differ from the nominal orbit, and 0 means that the error was too
large to even bother classifying the orbit.  The resonant objects are
shown with their resonance descriptor of the form M:N which refers to the
mean-motion resonance with Neptune, where M is the integer multiplier of
Neptune's period and N is the integer multiplier of the object's period.
Following the period relationship is listed the form and the order of
the resonance.  The symbol, $e$ or $i$ refers to the eccentricity or
inclination of the
object and $e_n$ refers to the eccentricity of Neptune.

\begin{table}
\caption{Summary of Object Types Found}\label{tbl6}
\begin{tabular*}{\maxfloatwidth}{lcccc} \hline
Orbit Type&
\multicolumn{1}{c}{Qual=3}&
\multicolumn{1}{c}{Qual=2}&
\multicolumn{1}{c}{Qual=1}&
\multicolumn{1}{c}{Qual=0} \\ \hline
1:1                    &    1&   0&   0& --- \\
2:1$e$                 &    3&   0&   0& --- \\
3:2$e$                 &   18&   6&   0& --- \\
3:2$e$+6:4$i^2$        &    1&   0&   0& --- \\
4:3$e$                 &    1&   0&   0& --- \\
5:2$e^3$               &    3&   1&   0& --- \\
5:3$e^2$               &    1&   1&   0& --- \\
5:4$e$                 &    1&   0&   0& --- \\
7:3$e^4$               &    0&   1&   0& --- \\
7:4$e^3$               &    4&   3&   0& --- \\
8:5$e^3$+8:5$e^2 e_n$  &    0&   1&   0& --- \\
9:5$e^4$               &    1&   1&   0& --- \\
Centaur                &    7&   2&   1& --- \\
Classical              &   70&   2&   0& --- \\
Scattered Disk         &   24&   3&   1& --- \\
Unknown                &  ---& ---& ---&  59 \\
\hline
\end{tabular*}
\end{table}

\section{Outer Edge of Belt}

The size of our sample is now large enough to address
the issue of the radial extent of the non-resonant (Classical) belt.
We can ask a very simple question of our data: is the outer limit of
our sample determined by the limits of our data collection or is it
determined by the spatial distribution of the belt?  The nominal
limiting magnitude for our data is $R=23.5$.  Within our dataset, 38\%
have a limiting magnitude fainter than 23.5, 44\% are between 23.0 and
23.5, 12\% are between 22.5 and 23.0, and the final 6\% are worse than
22.5.  These limiting magnitudes refer to the brighter limit of the pair
of frames used to search each location.  When the effects of followup
efforts are included, objects with good orbits have a limiting magnitude
of roughly $R=23$.  Given the range of $H_{\rm V}$ seen in the non-resonant
population, we should be able to see objects at a greater distance
than we do.  Additionally, as the heliocentric distance increases the
reflex motion of the object decreases.  However, our search strategy can
easily discern objects by their motion out to roughly 250 AU, perhaps
even further.  Our most distant discovery of any dynamical class
was an object at 53 AU.

\begin{figure}
\centerline{\includegraphics[width=12cm]{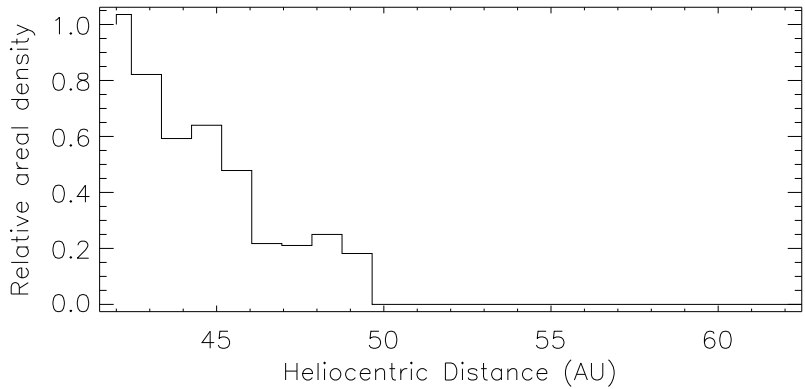}}
\caption{Number of non-resonant objects detected relative to the
number found at 42 AU.  This curve shows the observed fall off in sky density
of non-resonant objects as heliocentric distance increases.  If the
density and size-frequency-albedo distribution were independent of distance,
we would have measured 1 $\pm$ 0.4 at 55 AU.}\label{fig1}
\end{figure}

To see if this is a property of our survey or of the non-resonant belt,
we took all the non-resonant objects discovered between 41 and 43 AU.
Using this sample, we can ask how many of these would be detected at some
further distance.  In this way we can deduce a relative areal density as a
function of heliocentric distance.  This curve is shown in Fig.~\ref{fig1}.
Note that the observed relative areal density drops to zero at 50 AU.  Using
this
test, our survey would have seen objects out to 70 AU and this curve would be
flat if the areal density were independent of heliocentric distance.
The fact that this curve drops to zero is a direct indication of
a decrease in the number of objects with increasing distance.  This is the
so-called ``Kuiper Cliff'' which has previsouly been placed at 47 $\pm$
1 AU by \inlinecite{tru01}.  Our new data are clearly consistent with
this prior result.  We take Fig.~\ref{fig1} as solid evidence that the
space density or size-frequency-albedo distribution (or both) of non-resonant
objects changes dramatically with heliocentric distance.  Our sample
is sensitive out to 60-70 AU.  For reference, at 55 AU we should have
seen 7 objects compared to 28 objects in the observed sample.  But,
we did not find any classical object beyond 49 AU in our sample.

\section{Summary of Publically Available Resources}

The biggest challenge in reaching the goal of the survey is in obtaining
all the followup astrometry needed to secure the orbits of our newly
discovered objects.  A useful rule-of-thumb for the amount of astrometry
needed is 2-2-1, which translates to observations in 2 lunations in the
discovery apparition, observations in 2 lunations in the apparition
after discovery, and finally one more observation in one lunation
in the second year after discovery.  In this case, ``observation''
refers to collecting at least a pair of measurements on a single night.
This pattern of observation will usually determine the orbit well enough
to establish the dynamical type.  Looking at the problem this way, it
is clear that discovering objects is only 20\% of the work required.
It is also true that the power of our wonderful wide-field cameras is
wasted on most followup work.  We have had to increase the amount of time
we spend on followup to the detriment of new discoveries.  Only through
recovery observations with other telescopes will we reach the goal of
500 secure orbits.

Within the DES team we work aggressively to secure time on other
telescopes for followup on Mauna Kea, Lick Observatory, other telescopes
on Kitt Peak, Magellan and the Perkins 1.8-m.  Despite our best efforts,
objects are still being lost, but help from other observers can reduce the
loss rate.  To facilitate this community-wide collaborative effort we
make every effort to release all of our discoveries as soon as possible
following a search run.  These measurements are all submitted to the
Minor Planet Center (MPC), usually within 2-3 weeks after the end of
the run.  Additionally, we post a considerable amount of information
on the Lowell Observatory website.  Table~\ref{tbl7} summarizes some of
our online resources.  The information in these web pages
should be mostly self-explanatory.  However, take note that wherever an
object name appears, it is hyper-linked to a summary of information about
that object.  The summary includes the output from the orbit fitting
process \cite{ber00}, all of the astrometry used as input to the fit,
and the residuals from the fit.  Also note that there will often be
astrometry included in our data that are not published by the MPC.
We have included all data that we believe are relevant for the object.
Most often, the difference comes in partial recovery observations
where we have second apparition observations on only one night that
we believe constitute a good recovery.  However, these observations
are not considered an official recovery, and the measurements are not
made public by the MPC until another set of observations are collected.
These pages are regenerated automatically every morning.  It is our hope
that by providing these services and promptly reporting our new objects
that we enable others to assist with the formidable followup task.
We also encourage observers to send us a copy of any data submitted to
the MPC in case of any partial recovery observations.  If the linkages
appear to be valid, the observations will be added to our local database
and can be used immediately to guide future recovery efforts.

\begin{table}
\caption{Summary of Public Resources}\label{tbl7}
\begin{tabular*}{\maxfloatwidth}{lp{7.5cm}} \hline
URL& Content \\ \hline
http://www.lowell.edu& Main Lowell Observatory web site\\
\hskip10pt /Research/DES/& Main project summary page with links to other
information about the survey.\\
\hskip10pt /$\sim$buie/kbo& Main summary of DES objects and observing run
status.\\
\hskip20pt /kbofollowup.html& Table of links to followup lists.
The following links are currently
present but their content and organization may change in the future.\\
\hskip20pt /nondesig.html& List of our discoveries that are not yet
confirmed or designated and with
a positional error $< 1^{\circ}$.
This page may also indicate where the next followup attempt
will take place.\\
\hskip20pt /desig.html& List of all our discoveries that have been
designated regardless of their current ephemeris error.  This list is similar
to but not exactly the same as that generated by the Minor Planet
Center.  Regardless of formal discovery credit, an object appears here because
it can be included as part of the homogeneous dataset from the survey.\\
\hskip20pt /table1.html& List of all KBOs and Centaurs with little need of
astrometry.  This lists DES and non-DES objects.\\
\hskip20pt /table2.html& List of all KBOs and Centaurs needing astrometry but
whose errors are small.  This is a list of objects whose orbits would be
improved somewhat by new observations.  If you have a small-field instrument,
this list would be good to work from but are generally not critical to be
observed.\\
\hskip20pt /table3.html& List of all KBOs and Centaurs that need astrometry or
they will soon be lost.  This is the critical list of objects that need
observation.  Generally their errors are still small enough that they can be
found, but if they are not soon observed they will often be lost.\\
\hskip20pt /table4.html& List of all KBOs and Centaurs that have very large
positional errors and are essentially lost.  The ephemeris uncertainties on
these objects is generally large enough that a simple pointed recovery effort
will not be successful and the object must be re-discovered.\\
ftp://ftp.lowell.edu\\
\hskip5pt /pub/buie/kbo/recov\\
\hskip15pt /YYMMDD.dat& List of all designated KBOs and Centaurs at 0h UT on
the date
(YY--year, MM--month, and DD--day) given by the file name.  This file contains
predicted positions and uncertainties and other information about the
astrometry record.  This is designed to be used that night in support of
recovery observations.\\
\hskip15pt /YYMMDD.sdat& List of non-designated objects, same format as .dat
files (see {\tt tnorecov.pro} in IDL library).\\
\hskip5pt /pub/buie/idl& Repository of all IDL software used in this (and
other) projects.\\
\hline
\end{tabular*}
\end{table}

\section{Acknowledgements}

We thank help from numerous students, including K. Dekker, L. Hutchison, and
M. Trimble under the auspices of the NSF and MIT Undergraduate Research
Opportunities
Programs.  This research is based in part upon work supported by the NASA
Planetary Astronomy Progam through grants NAG5-8990, NAG5-10444, NAG5-13380,
and
NAG5-11058; STScI grant GO-9433; and the AAS.

\end{article}
\end{document}